# Strain-induced single-domain growth of epitaxial SrRuO$_3$ layers on SrTiO$_3$: a high-temperature x-ray diffraction study

Arturas Vailionis[1], Wolter Siemons[1,2], Gertjan Koster[1]

[1] Geballe Laboratory for Advanced Materials, Stanford University, Stanford, California, 94305, United States of America
[2] Faculty of Science and Technology and MESA+ Institute for Nanotechnology, University of Twente, 7500 AE, Enschede, The Netherlands

**Temperature dependent structural phase transitions of SrRuO$_3$ thin films epitaxially grown on SrTiO$_3$(001) single crystal substrates have been studied using high-resolution x-ray diffraction. In contrast to bulk SrRuO$_3$, coherently strained epitaxial layers do not display cubic symmetry up to ~730 °C and remain tetragonal. Such behavior is believed to be induced by compressive strain between the SrRuO$_3$ layer and SrTiO$_3$ substrate due to lattice mismatch. The tetragonal symmetry during growth explains the single domain growth on miscut SrTiO$_3$ substrates with step edges running along the [100] or [010] direction.**

The perovskite SrRuO$_3$ (SRO) thin films have attracted considerable interest due to their low room temperature resistivity and small lattice mismatch with a range of functional oxide materials.[1-3] In order for SrRuO$_3$ to grow coherently on a single crystal substrate, matching of layer in-plane lattice parameter to that of a substrate is required. The mismatch between in-plane lattice parameters of a film and the substrate introduces strain which affects the structural and electrical properties of the SrRuO$_3$ layer.

For bulk SrRuO$_3$, x-ray and neutron diffraction studies show that at room temperature the SrRuO$_3$ structure possesses an orthorhombic *Pbnm* symmetry with $a$ = 5.5670 Å, $b$ = 5.5304 Å, and $c$ = 7.8446 Å, similar to other ABO$_3$ perovskite compounds, and is isostructural with GdFeO$_3$.[4] The orthorhombic phase can be obtained by rotation of BO$_3$ (RuO$_3$) octahedra counterclockwise about the [010]$_{cubic}$ and [001]$_{cubic}$ directions and clockwise rotation about the [100]$_{cubic}$ direction of an ABO$_3$ cubic perovskite. At around 550 °C, the orthorhombic structure transforms into a tetragonal one with the space group *I4/mcm*.[5] In the tetragonal unit cell, RuO$_3$ octahedra are rotated only about the [001]$_{cubic}$ ABO$_3$ direction. At even higher temperatures of about 680 °C, tetragonal SrRuO$_3$ transforms into a cubic structure with a standard perovskite space group *Pm3m* where no rotations of RuO$_3$ octahedra are observed.[5]

As reported by Maria *et al.*,[6] thin SRO films grown on SrTiO$_3$(001) (STO) substrates undergo an orthorhombic to tetragonal (O-T) structural phase transition at a somewhat lower temperature of ~350 °C. They also measured a tetragonal to cubic (T-C) phase transition temperature of ~600 °C, but this was obtained from a bulk SrRuO$_3$ sample. The suggested transitions imply that SrRuO$_3$ exhibits cubic symmetry during film synthesis, which is typically in the range of 600-700 °C.[6] Since the T-C transition was observed on a powder SrRuO$_3$ sample, which does not represent conditions of commensurate strained layer growth, it is still unclear what crystal symmetry strained epitaxial SrRuO$_3$ layer possesses during growth under constrained geometries imposed by the substrate.

The only experiment that could clarify the presence of one or the other symmetry is to measure the structure of a SrRuO$_3$ layer at high temperatures, under the conditions that it is normally deposited on single crystal substrates. To see the T-C transition in a thin oriented SrRuO$_3$ film is not easy, because it involves just a slight rotation of the RuO$_3$ octahedra along the [001]$_{cubic}$ ABO$_3$ axis where only light oxygen atoms are involved. Using SrRuO$_3$ powder diffraction pattern simulations we established that the SrRuO$_3$(211) diffraction peak is very sensitive to oxygen rotation and is absent in the cubic symmetry. In this letter, capitalizing on this insight, we report a study of temperature-dependent structural transition of SrRuO$_3$ films coherently grown on SrTiO$_3$(001) substrates by high-resolution x-ray diffraction using laboratory as well as synchrotron radiation sources.

Thin film samples were grown by two different methods: Molecular Beam Epitaxy (MBE) and Pulsed Laser Deposition (PLD). The samples were grown in the same vacuum chamber with a background pressure of 10$^{-9}$ Torr. The MBE grown samples are deposited in an oxygen pressure of 10$^{-5}$ Torr and at a temperature of 700 °C. All films were grown on TiO$_2$ terminated SrTiO$_3$ substrates.[7] Samples were grown at a rate of about 1 Å/s from separate Ru and Sr sources. Typical thicknesses of the films range from 200 to 300 Å. During growth, Reflection High Energy Electron Diffraction (RHEED) is used to monitor the morphology of the samples. For PLD a 248 nm wavelength KrF excimer laser was employed with typical pulse lengths of 20-30 ns. The energy density on the target is kept at approximately 2.1 J/cm$^2$. Films were deposited with a laser repetition rate of 4 Hertz, with the substrate temperature at 700 °C.

X-ray diffraction (XRD) measurements were performed using a PANalytical X'Pert materials research diffractometer in high-resolution mode at the Stanford Nanocharacterization Laboratory as well as a synchrotron source at beamline 7-2 at the Stanford Synchrotron Radiation Laboratory (SSRL). For temperature dependent structural analysis an Anton-Paar hot stage was used.

The XRD results demonstrate that grown SrRuO$_3$ films exhibit (110) out-of-plane orientation with (100) and (-110) in-plane orientations along [100] and [010] directions of the STO substrate. Reciprocal lattice maps (RLM) taken at room temperature using symmetrical and asymmetrical reflections confirm that the SrRuO$_3$ layers are grown in a fully coherent fashion to the underlying SrTiO$_3$(001) substrate and exhibit a slightly distorted orthorhombic unit cell with the angle γ between the [100] and [010] directions being less that 90$^o$. Such a distortion was also reported by Gan et al.[8] Coherently grown SrRuO$_3$ is compressively strained along the [-110] and [001] directions. The in-plane compressive stress introduces out-of-plane strain that elongates the SrRuO$_3$ unit cell along the [110] direction. Such constrained in-plane geometry distorts the SrRuO$_3$ lattice from the ideal orthorhombic structure.

Figure 1 shows reciprocal lattice maps from the (260), (444), (620) and (44-4) reflections together with the SrTiO$_3$(204) reflections of a single domain SrRuO$_3$ layer. The difference in SrRuO$_3$ (260) and (620) atomic plane spacings represents dissimilarity in the *a* and *b* lattice parameters that is typical for the orthorhombic structure. In contrast, the tetragonal structure with *a* = *b* should show identical positions for the (260) and (620) Bragg reflections. In reciprocal space the transition from the orthorhombic to the tetragonal structure leads to an increase of the distance between SrRuO$_3$(260) and SrTiO$_3$(204) peaks while the distance between the SrRuO$_3$(620) and SrTiO$_3$(204) peaks will decrease.

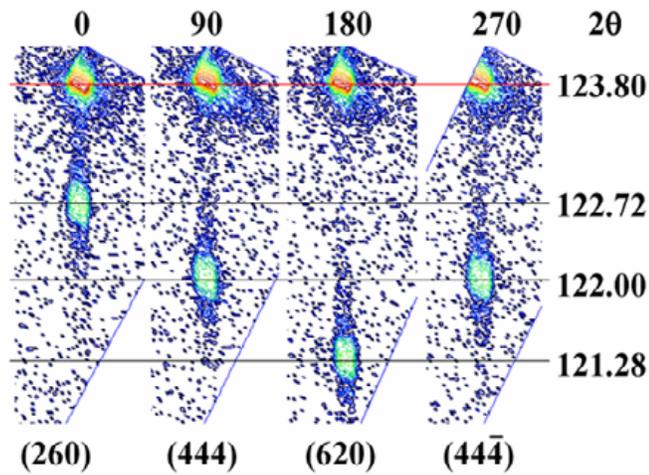

Fig. 1. Room temperature reciprocal space maps of $SrTiO_3(204)$ and $SrRuO_3(260)$, (444), (620), and (44-4) Bragg reflections. The fact that (260) and (620) peak positions are different from (444) and (44-4) positions indicates orthorhombic unit cell.

In order to examine the O-T transition, temperature-dependent XRD scans were performed along $l$-direction. As can be seen from Fig. 2, the distance between $SrTiO_3(204)$ and $SrRuO_3(620)$ peaks decreases as temperature increases. Above ~450 °C the $SrRuO_3(620)$ peak position reaches the same value as $SrRuO_3(260)$ completing the transition to the tetragonal structure. The inset of Fig. 2 shows the $SrRuO_3(221)$ peak intensity as a function of temperature. The disappearance of this peak at higher temperatures further proves the presence of a tetragonal phase with an O-T transition temperature of ~310 °C.

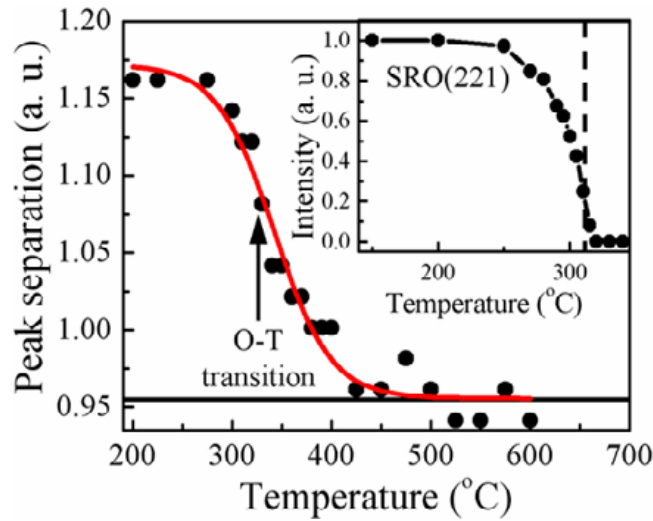

Fig. 2. The distance between $SrTiO_3(204)$ and $SrRuO_3(620)$ peaks in the reciprocal space as a function of temperature indicating orthorhombic to tetragonal transition. Inset shows the disappearance of $SrRuO_3(221)$ peak as film's symmetry changes to tetragonal. Black horizontal line indicates separation between $SrTiO_3(204)$ and $SrRuO_3(444)/(44-4)$ peak positions (see Fig. 1).

Finally, temperature dependent x-ray diffraction measurements of the SrRuO$_3$(211) diffraction peak were performed using synchrotron radiation. The SrRuO$_3$(211) peak intensity as a function of temperature is shown in Fig. 3. While the peak intensity gradually decreases indicating the rotation of the oxygen atoms, it does not vanish up to temperatures of ~730 $^o$C.

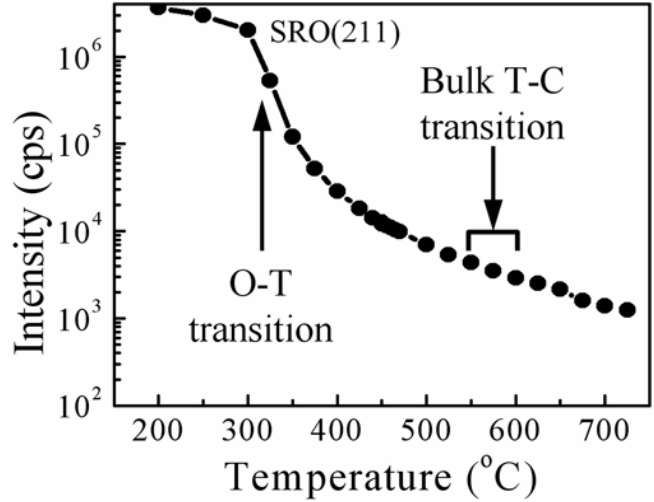

Fig. 3. Intensity of SrRuO$_3$(211) peak as a function of temperature. The nonzero intensity indicates that SrRuO$_3$ possesses tetragonal symmetry up to 730 $^o$C.

The results unambiguously demonstrate that a SrRuO$_3$(110) layer grown coherently on SrTiO$_3$(001) substrate does not undergo a tetragonal to cubic transition and remains tetragonal at temperatures up to 730 $^o$C. The tetragonal unit cell allows us to explain single domain growth of SrRuO$_3$ films on miscut SrTiO$_3$ substrates. In step flow growth mode, growing species tend to attach to the steps due to the larger diffusion length as compared to the terrace length of the substrate. For a tetragonal unit cell where $c > a = b$, SrRuO$_3$ will tend to align its *c*-axis along the steps. If the step edges run only along SrTiO$_3$ [100] or [010] directions then a single domain SrRuO$_3$ layer is formed. On the other hand, if the step edges run along direction rotated by some angle from [100] or [010] directions, SrRuO$_3$ will attach to steps with its longer unit cell axis parallel to the steps resulting in twinned structure, due to the serrated nature of the step edge, as was already observed by Gan *et al.*[9] SrRuO$_3$ layers on substrates with low miscut angles exhibit twinned structures due to large length of the substrate terraces as compared to the diffusion length of SrRuO$_3$, which results in island growth. In this regime SrRuO$_3$ tetragonal unit cell tends to align randomly along [100] and [010] directions of the SrTiO$_3$ substrate.

In summary, we have established the mechanism of single domain growth of epitaxial SrRuO$_3$ layers on single crystal SrTiO$_3$ substrates. We show that at room temperature SrRuO$_3$ films exhibit distorted orthorhombic structure with angle γ between [100] and [010] directions being less than 90$^o$. At a temperature of ~310 $^o$C, the SrRuO$_3$ layer undergoes a structural phase transition from orthorhombic to tetragonal. In contrast to the other studies, strained SrRuO$_3$ films do not show a structural phase transition from tetragonal to cubic and remain tetragonal up to temperatures of ~730 $^o$C.

This work was carried out under DoE BES support. Portions of this research were carried out at the Stanford Synchrotron Radiation Laboratory, a national user facility operated by Stanford University on behalf of the U.S. Department of Energy, Office of Basic Energy Sciences. The authors wish to acknowledge helpful discussion with J. Reiner, G. Rijnders, D.H.A. Blank, M.R. Beasley, T.H. Geballe and R.H. Hammond.

References
[1] B. Nagaraj, S. Aggarwal, and R. Ramesh, J. Appl. Phys. **90**, 375 (2001).
[2] G. Rijnders, D.H. A. Blank, J. Choi, and C.B. Eom, Appl. Phys. Lett. **84**, 505 (2004).
[3] J. Choi, C.B. Eom, G. Rijnders, H. Rogalla, and D.H. Blank, Appl. Phys. Lett. **79**, 1447 (2001).
[4] C.W. Jones, P.D. Battle, P. Lightfoot, and T.A. Harrison, Acta Cryst. C **45**, 365 (1989).
[5] B. J. Kennedy and B.A. Hunter, Phys. Rev. B **58**, 653 (1998).
[6] J. P. Maria, H.L. McKinstry, and S. Trolier-McKinstry, Appl. Phys. Lett. **76**, 3382 (2000).
[7] G. Koster, B.L. Kropman, G.J.H.M.Rijnders, D. Blank, and H. Rogalla, Appl. Phys. Lett. **73**, 2920 (1998).
[8] Q. Gan, R.A. Rao, C.B. Eom, L. Wu, and F. Tsui, J. Appl. Phys. **85**, 5297 (1999).
[9] Q. Gan, R.A. Rao, and C.B. Eom, Appl. Phys. Lett. **70**, 1962 (1997).